\documentclass[prl,aps,showpacs,twocolumn,floatfix]{revtex4-1}
\usepackage{amsmath,amsfonts,amssymb,bm,graphicx}
\newcommand{\bS}{{\bm S}}

\newcommand{\bsig}{{\bm\sigma}}
\newcommand{\e}{{\rm e}}
\newcommand{\ii}{{\rm i}}
\newcommand{\cd}{c^\dag}
\newcommand{\fd}{f^\dag}
\newcommand{\ua}{\uparrow}
\newcommand{\da}{\downarrow}

\begin{document}
\title{Heavy antiferromagnetic phases in Kondo lattices}
\author{L. Isaev}
\author{I. Vekhter}
\affiliation{Department of Physics and Astronomy, Louisiana State University,
             Baton Rouge LA 80703}
\begin{abstract}
 We propose a microscopic physical mechanism that stabilizes coexistence of the
 Kondo effect and antiferromagnetism in heavy-fermion systems. We consider a
 two-dimensional quantum Kondo-Heisenberg lattice model and show that
 long-range electron hopping leads to a robust antiferromagnetic Kondo state.
 By using a modified slave-boson mean-field approach we analyze the stability
 of the heavy antiferromagnetic phase across a range of parameters, and discuss
 transitions between different phases. Our results may be used to guide future
 experiments on heavy fermion compounds.
\end{abstract}
\pacs{71.10.Fd, 71.27.+a, 75.20.Hr}
\maketitle

\paragraph*{Introduction.}
The study of complex phenomena exhibited by materials with competing ground
states is the primary focus of modern condensed matter physics. In systems
where local magnetic moments interact with a conduction band there are two
opposing quantum many-body effects: Kondo screening (formation of singlets
between the local moments and itinerant electrons), and a long-range magnetic
[often antiferromagnetic (AF)] order. Since metallic Kondo phase and magnetic
ordering involve the same local-moment degrees of freedom, they contest the
same entropy. This competition is at the heart of the rich variety of phases
observed in heavy-fermion (HF) $f$-electron materials under tuning of external
parameters such as pressure, doping, or magnetic field \cite{coleman-2007}.

In the Kondo screened phase the Fermi surface (FS) volume accounts for the
local spins. Quantum (zero-temperature) phase transitions between this HF metal
with large FS and small-FS magnetic states remain an actively debated subject
\cite{coleman-2005,si-2010-1}. Of particular interest is whether the Kondo
screening collapses precisely at the onset of magnetism
\cite{coleman-2001,si-2010-2} or the magnetic transition is of a spin-density
wave (SDW) type \cite{hertz-1976,millis-1993} with no concomitant change of the
FS volume. Both scenarios are found in experiments \cite{si-2010-1}. In
${\rm Ce_3Pd_{20}Si_6}$ \cite{custers-2012} Kondo screening disappears inside a
magnetically ordered phase, while in ${\rm CeCu_2Si_2}$ the AF transition is
likely of the SDW type \cite{stockert-2008}. In ${\rm YbRh_2Si_2}$ data
indicate that Kondo screening disappears at the magnetic transition
\cite{paschen-2004,friedemann-2009}, while under Co and Ir doping the two
transitions separate \cite{friedemann-2009}. Developing microscopic theories
which exhibit such multitude of phases proved difficult
\cite{coleman-2007,si-2010-1,senthil-2004}. Recently it was conjectured
\cite{si-2010-1,coleman-2010} that frustrated magnetic interactions between
local moments give rise to a variety of phases that include magnetically
ordered as well as paramagnetic states with both large and small Fermi
surfaces, but detailed theories are still lacking.

In this Letter we propose a microscopic mechanism which controls the
coexistence of the Kondo effect and AF long-range order. We consider the
two-dimensional Kondo-Heisenberg lattice model with short-range AF interactions
between local moments, and conduction-electron hopping beyond nearest neighbors
(NN). By employing the notion of a ``spin-selective Kondo insulator''
\cite{peters-2012,li-2010} introduced in the context of ferromagnetism (FM) in
Kondo lattices we show that in a bipartite Kondo lattice with NN and
next-nearest neighbor (NNN) electron hopping the same physics leads to a robust
AF Kondo [``K+AF'' in Fig. \ref{fig2}(a)] phase. The stability of this state is
controlled by the relative magnitude of short- and long-range hopping
amplitudes. The AF Kondo phase has a large FS, and is separated by a
second-order quantum phase transition from the small-FS AF metal. In contrast
to previous studies \cite{zhang-2000,capponi-2001} we focus on the
experimentally relevant regime away from half-filling. We find the phases
somewhat similar to those suggested in Ref. \cite{senthil-2004}, but the
underlying physics is different, see discussion below.

\paragraph*{Model and approximations.}
The essential physics of magnetic HF materials is contained in the
Kondo-Heisenberg lattice model (KHLM),
\begin{align}
 H=&H_{\rm cond}+H_{\rm Kondo}+H_{\rm Heis}= \label{KLM_hamiltonian} \\
 &=-\sum_{(ij),\alpha}t_{ij}\bigl(\cd_{i\alpha}c_{j\alpha}+{\rm h.c.}\bigr)-
 \mu_c\sum_{i,\alpha}\cd_{i\alpha}c_{i\alpha}+ \nonumber \\
 &\qquad\qquad+J_K\sum_i\bS_i{\bm s}_i+J_H\sum_{\langle ij\rangle}\bS_i\bS_j.
 \nonumber
\end{align}
This Hamiltonian describes a system of itinerant electrons, $c_{i\alpha}$ with
spin indices $\alpha$, $\beta=\ua,\da$, interacting with local spin-$1/2$
moments $\bS_i$ via the AF Kondo coupling $J_K$. We consider this Hamiltonian
on a square lattice with sites labeled by $i$ and $j$. The competition to the
Kondo screening is provided by the AF exchange $J_H>0$ between NN spins. The
hopping amplitude $t_{ij}=t>0$ for NN links $(ij)=\langle ij\rangle$, and
$t_{ij}=t^\prime$ for NNN sites, $(ij)=\langle\langle ij\rangle\rangle$; for
all other sites $t_{ij}=0$. The electron spin operator is
${\bm s}_i=\bsig_{\alpha\beta}\cd_{i\alpha}c_{i\beta}/2$, where $\bsig$ are the
Pauli matrices. The chemical potential $\mu_c$ controls the conduction band
filling.

We employ the hybridization mean-field (HMF) approach \cite{coleman-2007},
where local spins are represented in terms of spin-$1/2$ pseudo-fermions,
$\bS_i=\bsig_{\alpha\beta}\fd_{i\alpha}f_{i\beta}/2$, subject to the constraint
$\sum_\alpha\fd_{i\alpha}f_{i\alpha}=1$, which removes unphysical empty and
doubly-occupied states from the single-site Hilbert space. The core idea of HMF
is to treat this constraint on the average by introducing the pseudo-fermion
``chemical potential'' $\mu_f$. The interactions in Eq. \eqref{KLM_hamiltonian}
now contain four fermion operators and are decoupled within the Hartree-Fock
approximation. We implement this approach along the lines of Ref.
\cite{kusminskiy-2008}, and consider decouplings in all possible channels,
namely:

(i) Magnetic channel
\begin{align}
 &H_{\rm Kondo}+H_{\rm Heis}\to \label{magnetic_ch} \\
 &\to J_K\sum_i\bigl({\bm M}_i{\bm s}_i+{\bm m}_i\bS_i\bigr)+J_H
 \sum_{\langle ij\rangle}\bigl({\bm M}_i\bS_j+{\bm M}_j\bS_i\bigr), \nonumber
\end{align}
where magnetic order parameters (OPs) are defined as
${\bm M}_i=\langle\bS_i\rangle$ and ${\bm m}_i=\langle{\bm s}_i\rangle$, and
the arrow indicates that we omitted c-number terms.

(ii) Pseudo-fermion dispersion (``spin-liquid'') channel
\begin{equation}
 H_{\rm Heis}\to-\frac{J_H}{4}\sum_{\langle ij\rangle}
 \bsig_{\alpha^\prime\alpha}\bsig_{\beta^\prime\beta}\bigl[\langle
 \fd_{i\alpha^\prime}f_{j\beta}\rangle\fd_{j\beta^\prime}f_{i\alpha}+{\rm h.c.}
 \bigr].
 \label{spin-liquid_ch}
\end{equation}

(iii) Kondo hybridization channel
\begin{align}
 H_{\rm Kondo}=&-\frac{3J_K}{4}\sum_i\chi^\dag_{i0}\chi_{i0}+\frac{J_K}{4}
 \sum_i{\bm\chi}^\dag_i{\bm\chi}_i\to \label{kondo_ch} \\
 &\to J_K\sum_i\biggl[-\frac{3}{4}\langle\chi_{i0}\rangle^*\chi_{i0}+
 \frac{1}{4}\langle{\bm\chi}_i\rangle^*{\bm\chi}_i+{\rm h.c.}\biggr] \nonumber
\end{align}
with $\chi_{i\mu}=\sigma^\mu_{\alpha\beta}\fd_{i\alpha}c_{i\beta}/\sqrt{2}$,
$\mu=0\ldots3$ and $\sigma^0_{\alpha\beta}=\delta_{\alpha\beta}$. Physically,
the $\chi$-operators are Schwinger bosons \cite{batista-2004} which create
local singlet and triplet states resulting from the Kondo coupling between
localized and itinerant spins. In the usual picture of Kondo singlet formation
only $\chi_0$-boson is present. The magnetic order admixes other components, so
that the $\chi$-representation of Eq. \eqref{kondo_ch} captures the $SU(2)$
invariance of the Kondo interaction \cite{kusminskiy-2008}.

Since we consider only uniform and commensurate AF states on the square
lattice, there are two wavevectors in the problem: ${\bm q}=0$ and
${\bm q}=Q_0=(\pi,\pi)$. In the AF phase all OPs are site-dependent. Thus
the pseudo-fermion density $n^f_i=\sum_\alpha\fd_{i\alpha}f_{i\alpha}$
will acquire an unphysical spatial dependence. To suppress the $Q_0$ harmonic
$n^f_{Q_0}$ we impose an additional constraint,
$H\!\!\to\!\!H\!-\!\mu_Q\sum_i\e^{\ii Q_0x_i}n^f_i$, where $\mu_Q$ is the
Lagrange multiplier.

\begin{figure}[t]
 \begin{center}
  \includegraphics[width=\columnwidth]{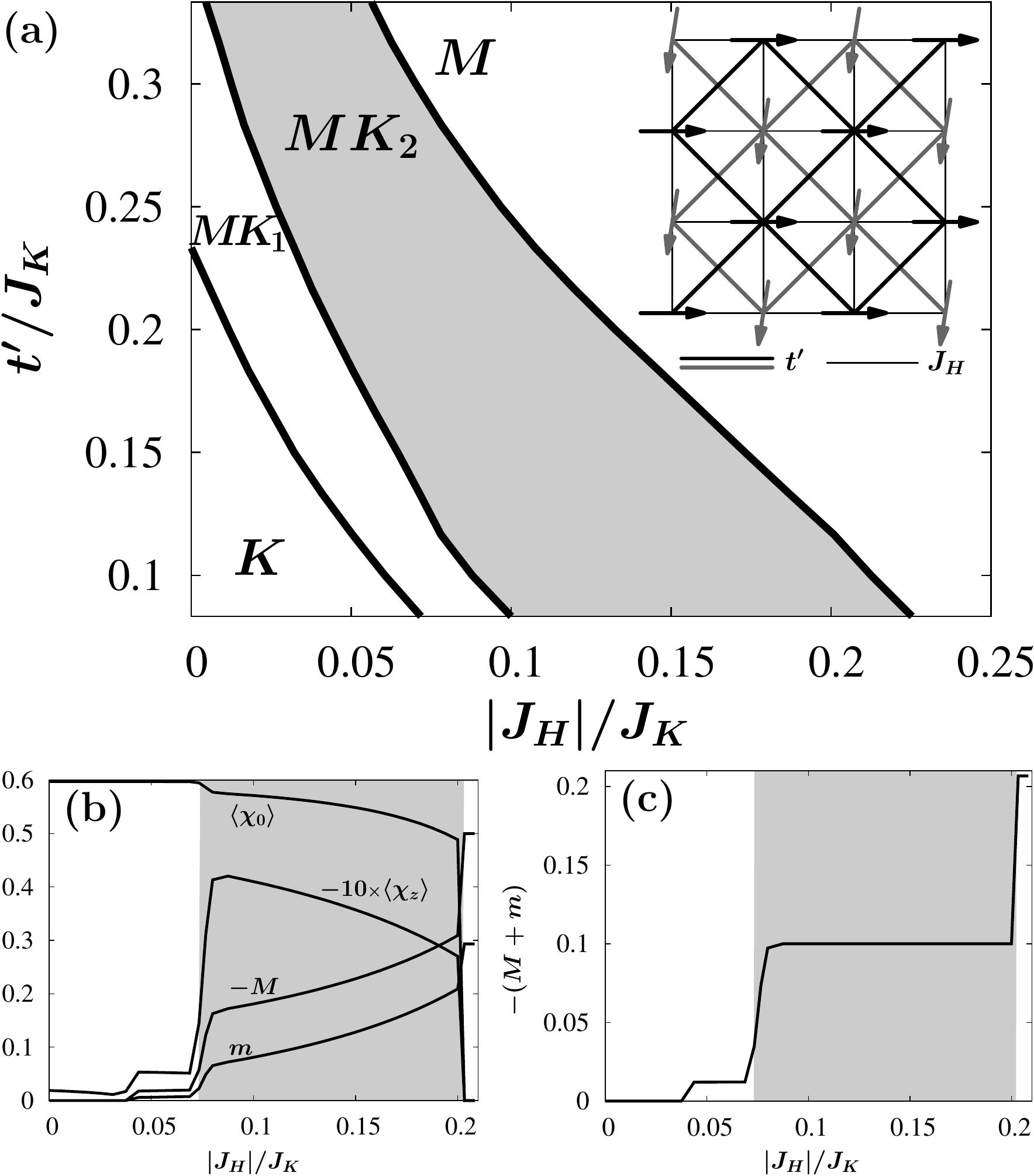}
 \end{center}
 \caption{KHLM with $t=0$. (a) $T=0$ phase diagram: ``K'' denotes singlet Kondo
	  phase, ``M'' is the local moment magnetic phase, ${\rm MK}_{1,2}$ are
	  magnetic Kondo phases. M and ${\rm MK}_{1,2}$ are ordered at
	  ${\bm q}=0$ (FM) if $J_H<0$ and $(\pi,\pi)$ (AF) for $J_H>0$. All
	  phase transitions are 1st order. Inset: Square lattice with $t=0$.
	  The arrows denote local moments. (b) and (c) Hybridization
	  $\langle\chi_{0,z}\rangle$, magnetic ($M$ and $m$) OPs, and total
	  magnetization $M+m$ for $t^\prime/J_K=2/15$. The plateau in (c)
	  signals that Eq. \eqref{commensurability_condition} is satisfied
	  inside ${\rm MK}_2$ but not ${\rm MK}_1$.}
 \label{fig1}
\end{figure}

In the rest of the paper we study the phase diagram of the KHLM as a function
of $t^\prime$, $J_H$ and temperature, $T$, at fixed density $n^c=0.8$. We
choose the spin quantization axis along the $z$-direction and omit the
corresponding vector indices whenever possible. At intermediate coupling we
need to sum over the entire Brillouin zone and numerically solve the HMF
equations in the momentum space on the $32\times32$ square lattice
with periodic boundary conditions, which is close to the thermodynamic limit.
Indeed increasing system size by a factor of 4 leads to only $\sim1\%$
correction to the results below.

\paragraph*{Origin of the AF Kondo phase: $t=0$ limit.}
This limit allows a simple analysis and guides subsequent discussion of
physical regimes. It is known \cite{peters-2012,li-2010} that the Kondo
lattice Hamiltonian with $J_H=0$ and NN hopping exhibits a ferromagnetic HF
state characterized by finite Kondo hybridization and magnetic polarization. In
that ``spin-selective Kondo insulator'' phase \cite{peters-2012}, away from
half-filling, all minority-spin and the corresponding fraction of the
majority-spin conduction electrons screen part of the $f$-electron spin, while
the magnetic moments of the remaining conduction and local $f$-electrons are
antiparallel to take advantage of the Kondo coupling.

To reveal the physical mechanism responsible for coexistence of the Kondo
effect and antiferromagnetism, it is instructive to start with a limiting case
with NN hopping $t=0$, but finite NNN hopping $t^\prime$. The model Eq.
\eqref{KLM_hamiltonian} then reduces to two interpenetrating square Kondo
(sub)lattices, coupled by the Heisenberg term [see inset in Fig.
\ref{fig1}(a)]. When $J_H=0$ and $t^\prime/J_K$ is large enough, each
sublattice enters a ferromagnetic Kondo state with non-zero
$\langle\chi_{0,z}\rangle$ and magnetic OPs. The ground state of this system is
continuously degenerate with respect to the angle between the magnetizations of
the two sublattices, similar to the $J_1$-$J_2$ antiferromagnet with NN ($J_1$)
and NNN ($J_2$) interactions for large $J_2/J_1$ \cite{Misguich}. Finite $J_H$
lifts this degeneracy and stabilizes long-range magnetic order coexisting with
Kondo singlets across the entire lattice. For $t=0$ only the Heisenberg
interaction couples the sublattices, and, at least at the level of HMF, the
ground state energy is an even function of $J_H$. The FM (AF) states are
stabilized for $J_H<0$ ($J_H>0$). Below, $m$ and $M$ denote uniform (for
$J_H<0$) and staggered (for $J_H>0$) $z$-axis magnetizations in the $c$- and
$f$-channels respectively. The vector part of the hybridization,
$\langle\chi_z\rangle$, has the same Fourier components as $m$ and $M$, while
the singlet hybridization amplitude $\langle\chi_0\rangle$ is always uniform.

Fig. \ref{fig1}(a) shows the $T=0$ mean field phase diagram of the KHLM with
$t=0$. For small $t^\prime$ and $\vert J_H\vert$ the system resides in the
Kondo singlet state. In the opposite limit, the Kondo effect is suppressed in
favor of magnetism. In the intermediate range of parameters several magnetic
Kondo states (${\rm MK}_{1,2}$) are stabilized. All phase transitions in Fig.
\ref{fig1}(a) are discontinuous as illustrated by the $\vert
J_H\vert$-dependence of the hybridization and magnetic OPs in Fig.
\ref{fig1}(b). The states ${\rm MK}_{1,2}$ can be distinguished based on the
commensurability condition \cite{peters-2012}
\begin{equation}
 2(M+m)=\vert1-n^c\vert,
 \label{commensurability_condition}
\end{equation}
satisfied only inside ${\rm MK}_2$. Consequently, the state ${\rm MK}_2$ has a
plateau in the total magnetization, shown in Fig. \ref{fig1}(c). We also note
that the phase ${\rm MK}_1$ is quite fragile and may be an artifact of the HMF
approximation.

\begin{figure}[t]
 \begin{center}
  \includegraphics[width=\columnwidth]{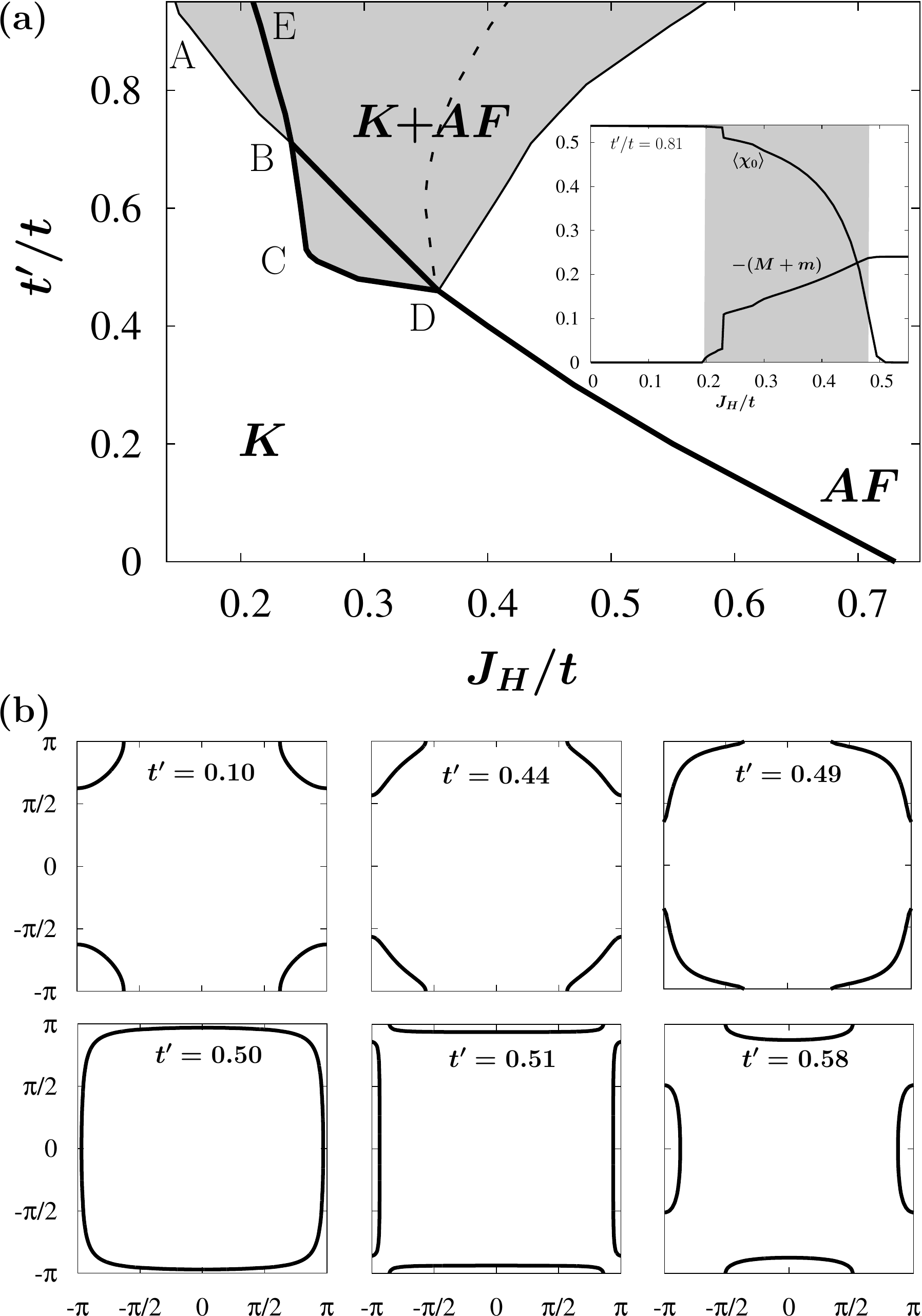}
 \end{center}
 \caption{KHLM at $T=0$. (a) Phase diagram. ``K'' is the singlet Kondo phase,
	  and ``${\rm K+AF}$'' is the large-FS AF Kondo state. The dashed line
	  is a continuation of the boundary between K and AF phases, obtained
	  if their coexistence is prohibited in the calculation. Thick (thin)
	  lines denote 1st (2nd) order transitions. Inset: Uniform part of the
	  Kondo hybridization OP and total staggered magnetization for
	  $t^\prime/t=0.81$. (b) Evolution of the heavy quasiparticle FS with
	  $t^\prime$ for $J_H=0$. The Lifshitz transition at
	  $t^\prime/t\sim0.5$ corresponds to point $C$ in panel (a).}
 \label{fig2}
\end{figure}

\paragraph*{Coexistence of antiferromagnetism and Kondo effect.}
The above picture survives in the physically relevant limit
$0\leqslant t^\prime/t\leqslant1$. Similar to the $t=0$ case, when $J_H=0$ and
$J_K$ are small compared to the electron bandwidth the system undergoes a
transition to a ferromagnetic Kondo state. To avoid this, we fix the Kondo
coupling at $J_K=6t$ (bandwidth is $\sim 8t$), so that for $J_H=0$ and any
$t^\prime\in[0,1]$ the ground state is the non-magnetic uniform HF phase.

Numerical solution of Eqs. \eqref{KLM_hamiltonian}--\eqref{kondo_ch} exhibits a
complex phase diagram shown in Fig. \ref{fig2}(a). Its most important feature
is the existence of a critical point {\it D} at $t^\prime_c/t\approx0.46$. For
$t^\prime<t^\prime_c$ there is a direct 1st order transition between the
large-FS HF and small-FS AF states. On the other hand, for
$t^\prime>t^\prime_c$ there is a large parameter range (shaded region in the
figure) where Kondo screening coexists with antiferromagnetism. The
hybridization OP is uniform in the Kondo phase,
$\langle\chi_{i0}\rangle=\langle\chi_0\rangle$, and vanishes in the AF state.
The latter is characterized by $M^z_i=M\e^{\ii Q_0{\bm x}_i}$ and
$m^z_i=m\e^{\ii Q_0{\bm x}_i}$. In the coexistence phase (${\rm K+AF}$) the
singlet component of the hybridization acquires a small staggered component,
$\langle\chi_{i0}\rangle=\langle\chi_0\rangle+\langle{\tilde\chi}_0\rangle
\e^{\ii Q_0{\bm x}_i}$
($\vert\langle{\tilde\chi}_0\rangle/\langle\chi_0\rangle\vert\sim0.01$), while
the vector part of the Kondo hybridization is purely staggered,
$\langle\chi_{iz}\rangle=\langle\chi_z\rangle\e^{\ii Q_0{\bm x}_i}$, tracking
the spatial variations of the magnetization.

The coexistence phase is separated from the AF state by a 2nd order transition.
However, its boundary with the pure Kondo state is more complex. In Fig.
\ref{fig2}(a) the line $AB$ is the 2nd order transition, line $DCB$ is weakly
1st order, and $BD$ and $BE$ are strongly 1st order transitions. The AF Kondo
state inside regions $ABE$ and $BCD$ differs from the rest of the intermediate
phase only in its value of the staggered magnetization $M+m$, see inset of
Fig. \ref{fig2}(a). Absence of the magnetization plateau in this phase [cf.
Fig. \ref{fig1}(c)] indicates that the commensurability condition
\eqref{commensurability_condition} no longer holds in the presence of the NN
hopping.

\begin{figure}[t]
 \begin{center}
  \includegraphics[width=\columnwidth]{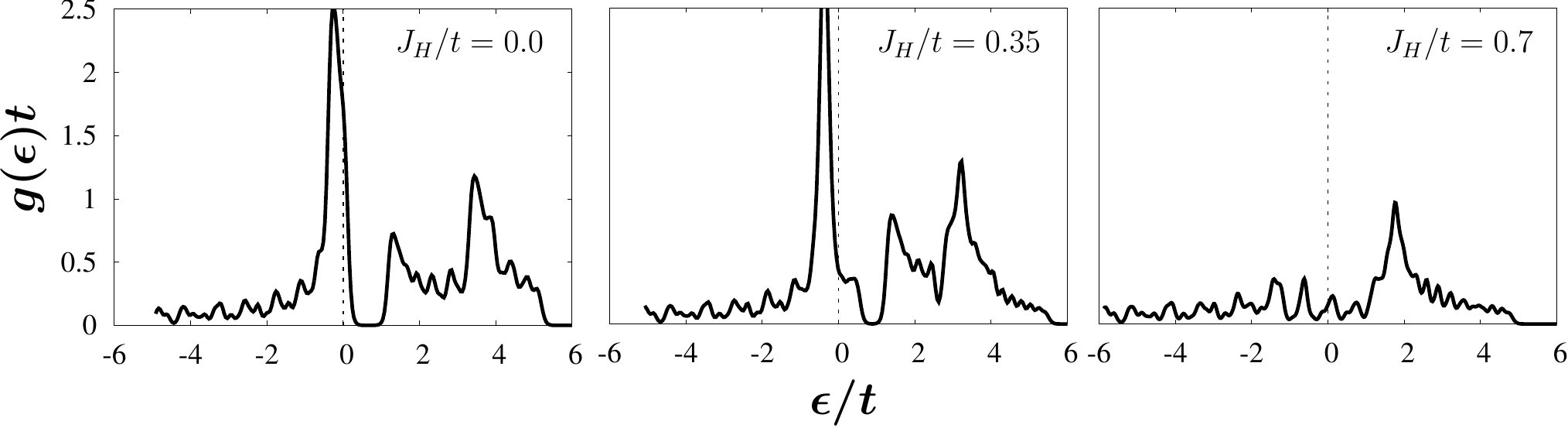}
 \end{center}
 \caption{Quasiparticle DOS for a fixed $t^\prime/t=0.81$ and $J_H$ inside the
	  singlet Kondo phase (left), magnetic Kondo state (middle) and AF
	  metal (right). The energy $\epsilon$ is measured relative to the
	  Fermi level. In the left and middle panels the Kondo peak is
	  apparent.}
 \label{fig3}
\end{figure}

The kink in the phase boundary at point $C$ is due to the change in the
topology of the heavy quasiparticle FS (Lifshitz transition) at
$t^\prime/t\sim0.5$ in Fig. \ref{fig2}(b) \cite{c-comment}. Hence, the
intermediate phases $ABE$ and $BCD$ may signify tendency towards incommensurate
ordering, rather than the $(\pi,\pi)$ AF state that we consider. More studies
are needed to determine the exact spatial structure of the coexistence state.
Although the AF Kondo state exists only because of NNN hopping, increasing
$t^\prime$ eventually destroys it when $J_K$ becomes small compared to the
bandwidth.

Finally, in Fig. \ref{fig3} we present the quasiparticle density of states
(DOS) in the three phases along the line $t^\prime/t=0.81$ in Fig.
\ref{fig2}(a). As expected, the DOS in the AF Kondo phase shows a peak near the
Fermi surface, similar to the singlet Kondo state, and is very different from
the DOS in the small-FS AF metal.

\begin{figure}[t]
 \begin{center}
  \includegraphics[width=\columnwidth]{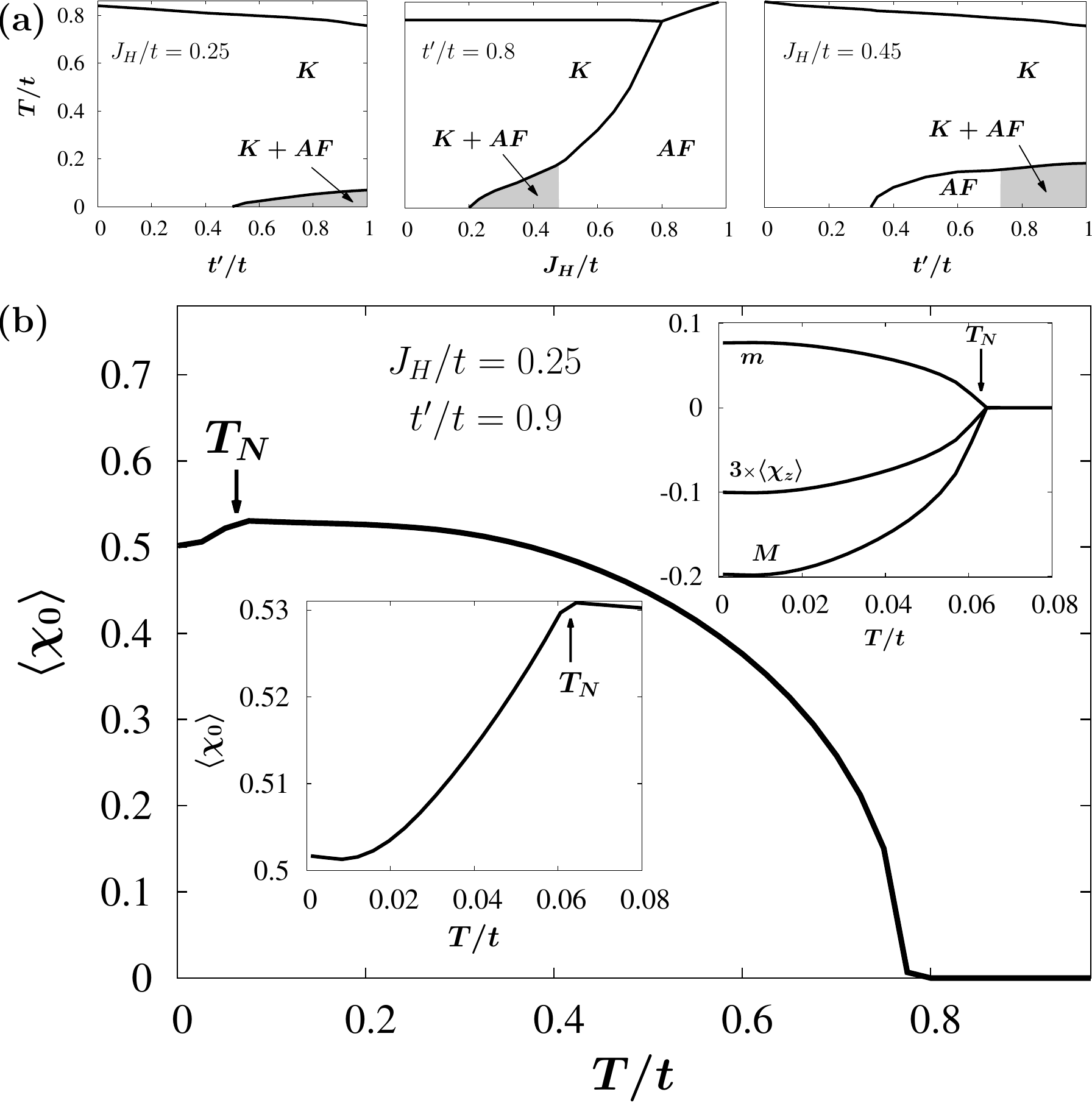}
 \end{center}
 \caption{KHLM at finite $T$. (a) Phase diagrams exhibiting competing orders,
          see text. Notations are the same as in Fig. \ref{fig2}. Shaded
          regions denote AF Kondo phases. The transition between Kondo and AF
          (AF Kondo) states is 1st (2nd) order. Unmarked phases are usual Fermi
          liquids. (b) OPs in the AF Kondo phase for a fixed $J_H/t=0.25$ and
          $t^\prime/t=0.9$: $\langle\chi_0\rangle$ (main panel and lower
          inset), and $\langle\chi_z\rangle$ and staggered magnetizations
          (upper inset).}
 \label{fig4}
\end{figure}

\paragraph*{Finite temperature behavior.}
To assess thermal stability of the AF Kondo state, in Fig. \ref{fig4}(a) we
present the finite-$T$ phase diagram of the KHLM computed along the
$\prod$-shaped path centered at point $D$ of Fig. \ref{fig2}(a). In the HMF
analysis the Kondo screened phase ($K$) always vanishes via a 2nd
order phase transition to a small-FS metal at a critical temperature
$T_K(J_H,t^\prime)$. Depending on $J_H$ and $t^\prime$ the HF metal may become
unstable towards either a pure AF state or a ``${\rm K+AF}$'' phase [shaded
regions in Fig. \ref{fig4}(a)] at a N\'eel temperature $T_N(J_H,t^\prime)<T_c$.
The latter phase extends over a sizable range $T_N/T_c\lesssim 1/4$ and is
separated from the singlet Kondo state by a 2nd order transition. This is
illustrated in Fig. \ref{fig4}(b) which shows the temperature dependence of all
OPs. In contrast, pure AF phase is separated from the HF state by a 1st order
transition.

It is interesting to observe that Fig. \ref{fig4} resembles the phase diagram
of a spin-$1$ underscreened KLM \cite{perkins-2007} if $T_N$ is replaced by the
Curie temperature. Finally, we note that, although spontaneous continuous
symmetry-breaking in a pure 2D system is prohibited by the Mermin-Wagner
theorem \cite{mermin-1966}, an infinitesimal coupling to the third dimension
will be sufficient to stabilize the ordered phases in Fig. \ref{fig4}.

\paragraph*{Discussion.}
Our study identifies the long-range electron hopping as a physical mechanism
responsible for the robust coexistence of antiferromagnetism and Kondo effect
in the description of HF materials via the Kondo-Heisenberg lattice model. The
parameter $t^\prime/t$ controls whether the Kondo screening can vanish
precisely at the onset of the magnetic order, or via and intermediate
coexistence regime, covering the range of experimental data. Although we
considered a square lattice, our results can be straightforwardly applied to
any bipartite graph.

The phase diagram in Fig. \ref{fig2}(a) can be tested in HF systems by changing
$t^\prime/t$ with pressure. However, as $J_H$ is also usually
pressure-sensitive, the exact path cut in Fig. \ref{fig2}(a) by such an
experiment is not clear and either an AF or ${\rm K+AF}$ phase may be realized.
Chemical pressure, such as substitution of ${\rm Ni}$ for ${\rm Pd}$, or
${\rm Co}$ for ${\rm Rh}$ in ${\rm Ce}$- or ${\rm Yb}$-based HF compounds, can
also be used to explore the phases and the transitions between them. Study of
the critical behavior near point $D$ would be particularly intriguing.

We emphasize that our study is distinct from previous works, notably Ref.
\onlinecite{senthil-2004} which considered the model \eqref{KLM_hamiltonian}
with $t^\prime\equiv 0$ and found a phase diagram that included the heavy SDW
state similar to our AF Kondo phase. However, their analysis involved enforcing
the particular form of the spin-liquid order and imposing unequal coupling
constants in Eqs. \eqref{magnetic_ch} and \eqref{spin-liquid_ch} to stabilize
it (in our case they are both equal to $J_H$). We provided a framework where
the fully self-consistent treatment of the KHLM allows controlled analysis to
the different OPs. Crucially, while we find the non-zero ``spin-liquid'' order
parameter in all the Kondo screened phases, it vanishes in the local moment
antiferromagnet, and therefore the spinon FS is absent in that phase.

It is known that the finite-$T$ HMF phase transition between HF and small-FS
metal phases becomes a crossover when fluctuations beyond HMF are taken into
account. However, the condensation of bosons $\chi_0$ and $\bm\chi$ does remain
a phase transition at $T=0$ \cite{senthil-2004}, and hence salient features of
the quantum phase diagram of Fig. \ref{fig2}(a) remain unchanged, although
phase transition lines may shift. At finite $T$ we expect a wide
quantum critical regime above those transition lines.

We focused on commensurate magnetic phases. However, at least in
the classical KHLM such states are often suppressed in favor of incommensurate
spiral \cite{hamada-1995} or skyrmion \cite{solenov-2012} phases. Therefore it
would be interesting to establish whether such states are realized in the {\it
quantum} KHLM. We leave this for future investigations.

We acknowledge support by DOE Grant DE-FG02-08ER46492 (L.I.) and NSF
Grant DMR-1105339 (I.V.).


\begin{thebibliography}{99}
\bibitem{c-comment}
  It is possibile that around point $C$ there exists a narrow AF Kondo phase.
  However, mapping its boundaries is beyond the accuracy of the HMF
  approximation.
\end{thebibliography}
\end{document}